# Customer churn prediction in telecom using machine learning in big data platform

Abdelrahim Kasem Ahmad*, Assef Jafar and Kadan Aljoumaa

*Correspondence: Abdelrahim.ahmad@hiast.edu.sy
Faculty of Information Technology, Higher Institute for Applied Sciences and Technology, Damascus, Syria

**Abstract**

Customer churn is a major problem and one of the most important concerns for large companies. Due to the direct effect on the revenues of the companies, especially in the telecom field, companies are seeking to develop means to predict potential customer to churn. Therefore, finding factors that increase customer churn is important to take necessary actions to reduce this churn. The main contribution of our work is to develop a churn prediction model which assists telecom operators to predict customers who are most likely subject to churn. The model developed in this work uses machine learning techniques on big data platform and builds a new way of features' engineering and selection. In order to measure the performance of the model, the Area Under Curve (AUC) standard measure is adopted, and the AUC value obtained is 93.3%. Another main contribution is to use customer social network in the prediction model by extracting Social Network Analysis (SNA) features. The use of SNA enhanced the performance of the model from 84 to 93.3% against AUC standard. The model was prepared and tested through Spark environment by working on a large dataset created by transforming big raw data provided by SyriaTel telecom company. The dataset contained all customers' information over 9 months, and was used to train, test, and evaluate the system at SyriaTel. The model experimented four algorithms: Decision Tree, Random Forest, Gradient Boosted Machine Tree "GBM" and Extreme Gradient Boosting "XGBOOST". However, the best results were obtained by applying XGBOOST algorithm. This algorithm was used for classification in this churn predictive model.

## Introduction

The telecommunications sector has become one of the main industries in developed countries. The technical progress and the increasing number of operators raised the level of competition [1]. Companies are working hard to survive in this competitive market depending on multiple strategies. Three main strategies have been proposed to generate more revenues [2]: (1) acquire new customers, (2) upsell the existing customers, and (3) increase the retention period of customers. However, comparing these strategies taking the value of return on investment (RoI) of each into account has shown that the third strategy is the most profitable strategy [2], proves that retaining an existing customer costs much lower than acquiring a new one [3], in addition to being considered much easier than the upselling strategy [4]. To apply the third strategy, companies have to decrease the potential of customer's churn, known as "the customer movement from one provider to another" [5].

Customers' churn is a considerable concern in service sectors with high competitive services. On the other hand, predicting the customers who are likely to leave the company will represent potentially large additional revenue source if it is done in the early phase [3].

Many research confirmed that machine learning technology is highly efficient to predict this situation. This technique is applied through learning from previous data [6, 7].

The data used in this research contains all customers' information throughout nine months before baseline. The volume of this dataset is about 70 Terabyte on HDFS "Hadoop Distributed File System", and has different data formats which are structured, semi-structured, and unstructured. The data also comes very fast and needs a suitable big data platform to handle it. The dataset is aggregated to extract features for each customer.

We built the social network of all the customers and calculated features like degree centrality measures, similarity values, and customer's network connectivity for each customer. SNA features made good enhancement in AUC results and that is due to the contribution of these features in giving more different information about the customers.

We focused on evaluating and analyzing the performance of a set of tree-based machine learning methods and algorithms for predicting churn in telecommunications companies. We have experimented a number of algorithms such as Decision Tree, Random Forest, Gradient Boost Machine Tree and XGBoost tree to build the predictive model of customer Churn after developing our data preparation, feature engineering, and feature selection methods.

There are two telecom companies in Syria which are SyriaTel and MTN. SyriaTel company was interested in this field of study because acquiring a new customer costs six times higher than the cost of retaining the customer likely to churn. The dataset provided by SyriaTel had many challenges, one of them was unbalance challenge, where the churn customers' class was very small compared to the active customers' class. We experimented three scenarios to deal with the unbalance problem which are oversampling, undersampling and without re-balancing. The evaluation was performed using the Area under receiver operating characteristic curve "AUC" because it is generic and used in case of unbalanced datasets [8].

Many previous attempts using the Data Warehouse system to decrease the churn rate in SyriaTel were applied. The Data Warehouse aggregated some kind of telecom data like billing data, Calls/SMS/Internet, and complaints. Data Mining techniques were applied on top of the Data Warehouse system, but the model failed to give high results using this data. In contrast, the data sources that are huge in size were ignored due to the complexity in dealing with them. The Data Warehouse was not able to acquire, store, and process that huge amount of data at the same time. In addition, the data sources were from different types, and gathering them in Data Warehouse was a very hard process so that adding new features for Data Mining algorithms required a long time, high processing power, and more storage capacity. On the other hand, all these difficult processes in Data Warehouse are done easily using distributed processing provided by big data platform.

Furthermore, big social networks, as those in SyriaTel, are considered one of the fundamental components of big data network graphs [9]. The computational complexity of SNA measures is very high due to the nature of the iterative calculations done on a big scale graph, as mentioned in Eqs. (1) and (2). A lot of work to decrease the complexity of computing SNA measures has been done. For example, Barthelemy [10] proposed a new algorithm to reduce the complexity of calculating the Betweenness centrality from O(n3) to O(n2). Elisabetta [11] also proposed an approximation method to compute the Betweenness with less complexity. In spite of that, the traditional Data Warehouse system still suffers from deficiencies in computing the essential SNA measures on large scale networks.

Big data system allowed SyriaTel Company to collect, store, process, aggregate the data easily regardless of its volume, variety, and complexity. In addition, it enabled extracting richer and more diverse features like SNA features that provide additional information to enhance the churn predictive model.

We believe that big data facilitated the process of feature engineering which is one of the most difficult and complex processes in building predictive models. By using the big data platform, we give the power to SyriaTel company to go farther with big data sources. In addition, the company becomes able to extract the Social Network Analysis features from a big scale social graph which is built from billions of edges (transactions) that connect millions of nodes (customers). The hardware and the design of the big data platform illustrated in "Proposed churn method" section fit the need to compute these features regardless of their complexity on this big scale graph.

The model also was evaluated using a new dataset and the impact of this system to the decision to churn was tested. The model gave good results and was deployed to production.

## Related work

Many approaches were applied to predict churn in telecom companies. Most of these approaches have used machine learning and data mining. The majority of related work focused on applying only one method of data mining to extract knowledge, and the others focused on comparing several strategies to predict churn.

Gavril et al. [12] presented an advanced methodology of data mining to predict churn for prepaid customers using dataset for call details of 3333 customers with 21 features, and a dependent churn parameter with two values: Yes/No. Some features include information about the number of incoming and outgoing messages and voicemail for each customer. The author applied principal component analysis algorithm "PCA" to reduce data dimensions. Three machine learning algorithms were used: Neural Networks, Support Vector Machine, and Bayes Networks to predict churn factor. The author used AUC to measure the performance of the algorithms. The AUC values were 99.10%, 99.55% and 99.70% for Bayes Networks, Neural networks and support vector machine, respectively. The dataset used in this study is small and no missing values existed.

He et al. [13] proposed a model for prediction based on the Neural Network algorithm in order to solve the problem of customer churn in a large Chinese telecom company which contains about 5.23 million customers. The prediction accuracy standard was the overall accuracy rate, and reached 91.1%.

Idris [14] proposed an approach based on genetic programming with AdaBoost to model the churn problem in telecommunications. The model was tested on two standard data sets. One by Orange Telecom and the other by cell2cell, with 89% accuracy for the cell2cell dataset and 63% for the other one.

Huang et al. [15] studied the problem of customer churn in the big data platform. The goal of the researchers was to prove that big data greatly enhance the process of predicting the churn depending on the volume, variety, and velocity of the data. Dealing with data from the Operation Support department and Business Support department at China's largest telecommunications company needed a big data platform to engineer the fractures. Random Forest algorithm was used and evaluated using AUC.

Makhtar et al. [16] proposed a model for churn prediction using rough set theory in telecom. As mentioned in this paper Rough Set classification algorithm outperformed the other algorithms like Linear Regression, Decision Tree, and Voted Perception Neural Network.

Various researches studied the problem of unbalanced data sets where the churned customer classes are smaller than the active customer classes, as it is a major issue in churn prediction problem. Amin et al. [17] compared six different sampling techniques for oversampling regarding telecom churn prediction problem. The results showed that the algorithms (MTDF and rules-generation based on genetic algorithms) outperformed the other compared oversampling algorithms.

Burez and Van den Poel [8] studied the problem of unbalance datasets in churn prediction models and compared performance of Random Sampling, Advanced Under-Sampling, Gradient Boosting Model, and Weighted Random Forests. They used (AUC, Lift) metrics to evaluate the model. the result showed that undersampling technique outperformed the other tested techniques.

We did not find any research interested in this problem recorded in any telecommunication company in Syria. Most of the previous research papers did not perform the feature engineering phase or build features from raw data while they relied on ready features provided either by telecom companies or published on the internet.

In this paper, the feature engineering phase is taken into consideration to create our own features to be used in machine learning algorithms. We prepared the data using a big data platform and compared the results of four trees based machine learning algorithms.

## Data set

There are many types of data in SyriaTel used to build the churn model. These types are classified as follow:

1. *Customer data* It contains all data related to customer's services and contract information. In addition to all offers, packages, and services subscribed to by the customer. Furthermore, it also contains information generated from CRM system like (all customer GSMs, Type of subscription, birthday, gender, the location of living and more ...).

2. *Towers and complaints database* The information of action location is represented as digits. Mapping these digits with towers' database provides the location of this transaction, giving the longitude and latitude, sub-area, area, city, and state.

   Complaints' database provides all complaints submitted and statistics inquiries related to coverage, problems in offers and packages, and any problem related to the telecom business.
3. *Network logs data* Contains the internal sessions related to internet, calls, and SMS for each transaction in Telecom operator, like the time needed to open a session for the internet and call ending status. It could indicate if the session dropped due to an error in the internal network.
4. *Call details records "CDRs"* Contain all charging information about calls, SMS, MMS, and internet transaction made by customers. This data source is generated as text files.
5. *Mobile IMEI information* It contains the brand, model, type of the mobile phone and if it's dual or mono SIM device.

   This data has a large size and there is a lot of detailed information about it. We spent a lot of time to understand it and to know its sources and storing format. In addition to these records, the data must be linked to the detailed data stored in relational databases that contain detailed information about the customer. The nine months of data sets contained about ten million customers. The total number of columns is about ten thousand columns.

## Data exploration and challenges with SyriaTel dataset

Spark engine is used to explore the structure of this dataset, it was necessary to make the exploration phase and make the necessary pre-preparation so that the dataset becomes suitable for classification algorithms. After exploring the data, we found that about 50% of all numeric variables contain one or two discrete values, and nearly 80% of all the categorical variables have Less than 10 categories, 15% of the numerical variables and 33% of the categorical variables have only one value. Most of some variables' values are around zero. We found that 77% of the numerical variables have more than 97% of their values filled with 0 or null value. These results indicate that a large number of variables can be removed because these variables are fixed or close to a constant. This dataset encounters many challenges as follow.

### Data volume

Since we don't know the features that could be useful to predict the churn, we had to work on all the data that reflect the customer behavior in general. We used data sets related to calls, SMS, MMS, and the internet with all related information like complaints, network data, IMEI, charging, and other. The data contained transactions for all customers during nine months before the prediction baseline. The size of this data was more than 70 Terabyte, and we couldn't perform the needed feature engineering phase using traditional databases.

### Data variety

The data used in this research is collected from multiple systems and databases. Each source generates the data in a different type of files as structured, semi-structured (XML-JSON) or unstructured (CSV-Text). Dealing with these kinds of data types is very hard without big data platform since we can work on all the previous data types without making any modification or transformation. By using the big data platform, we no longer have any problem with the size of these data or the format in which the data are represented.

### Unbalanced dataset

The generated dataset was unbalanced since it is a special case of the classification problem where the distribution of a class is not usually homogeneous with other classes. The dominant class is called the basic class, and the other is called the secondary class. The data set is unbalanced if one of its categories is 10% or less compared to the other one [18].

Although machine learning algorithms are usually designed to improve accuracy by reducing error, not all of them take into account the class balance, and that may give bad results [18]. In general, classes are considered to be balanced in order to be given the same importance in training.

We found that SyriaTel dataset was unbalanced since the percentage of the secondary class that represents churn customers is about 5% of the whole dataset.

### Extensive features

The collected data was full of columns, since there is a column for each service, product, and offer related to calls, SMS, MMS, and internet, in addition to columns related to personnel and demographic information. If we need to use all these data sources the number of columns for each customer before the data being processed will exceed ten thousand columns.

### Missing values

There is a representation of each service and product for each customer. Missing values may occur because not all customers have the same subscription. Some of them may have a number of services and others may have something different. In addition, there are some columns related to system configurations and these columns have only null value for all customers.

## Proposed churn method

In order to build the churn predictive system at SyriaTl, a big data platform must be installed. Hortonworks Data Platform (HDP)[1] was chosen because it is a free and an open source framework. In addition, it is under the Apache 2.0 License. HDP platform has a variety of open source systems and tools related to big data. These open source systems and tools are integrated with each other. Figure 1 presents the ecosystem of HDP, where each group of tools is categorized under specific specialization like Data Management, Data Access, Security, Operations and Governance Integration.

[1] https://hortonworks.com/.

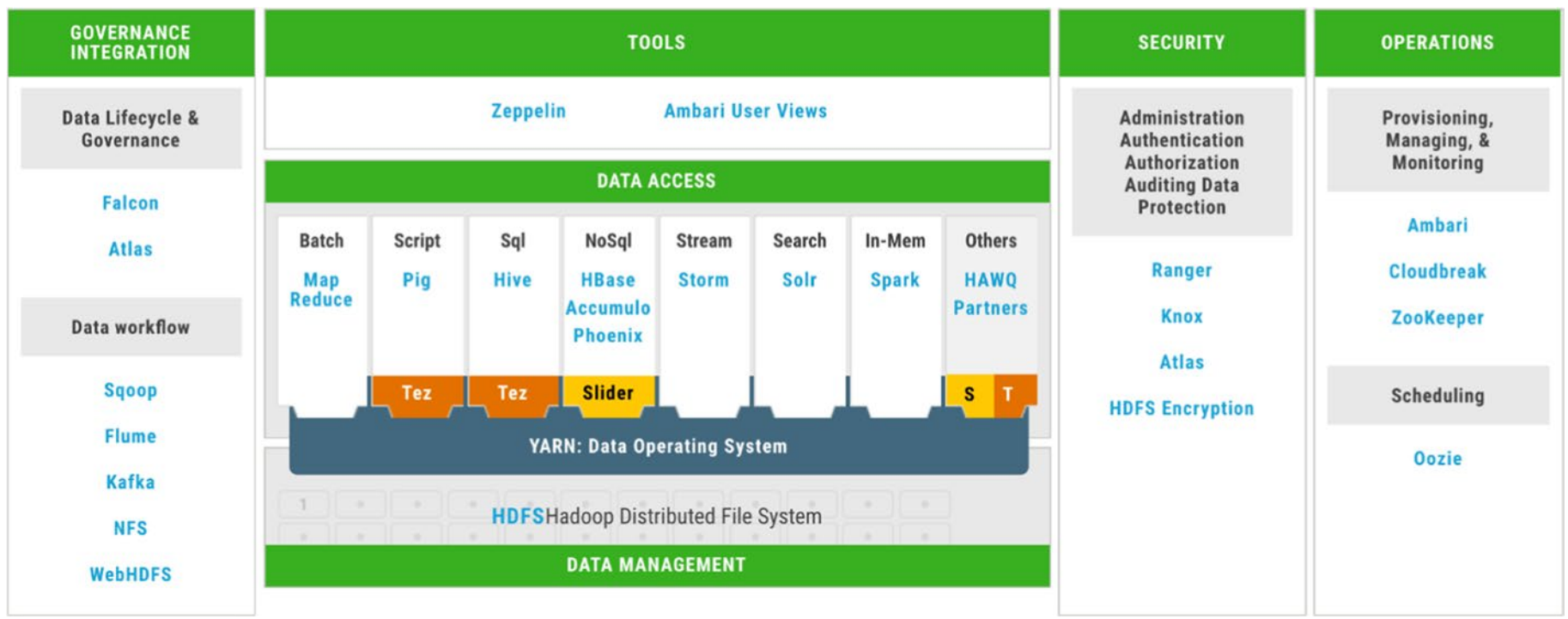

**Fig. 1** Hortonworks data platform HDP—big data framework

The installation of HDP framework was customized in order to have the only needed tools and systems that are enough to go through all phases of this work. This customized package of installed systems and tools is called SYTL-BD framework (SyriaTel's big data framework). We installed Hadoop Distributed File System HDFS[2] to store the data, Spark execution engine[3] to process the data, Yarn[4] to manage the resources, Zeppelin[5] as the development user interface, Ambari[6] to monitor the system, Ranger[7] to secure the system and (Flume[8] System and Scoop[9] tool) to acquire the data from outside SYTL-BD framework into HDFS.

The used hardware resources contained 12 nodes with 32 Gigabyte RAM, 10 Terabyte storage capacity, and 16 cores processor for each node. A nine consecutive months dataset was collected. This dataset will be used to extract the features of churn predictive model. The data life cycle went through several stages as shown in Fig. 2

Spark engine was used in most of the phases of the model like data processing, feature engineering, training and testing the model since it performs the processing on RAM. In addition, there are many other advantages. One of these advantages is that this engine containing a variety of libraries for implementing all stages of machine learning lifecycle.

## Data acquisition and storing

Moving the data from outside SYTL-BD into HDFS was the first step of work. The data is divided into three main types which are structured, semi-structured and unstructured.

Apache Flume is a distributed system used to collect and move the unstructured (CSV and text) and semi-structured (JSON and XML) data files to HDFS. Figure 3 shows

2 https://hadoop.apache.org/docs/r1.2.1/hdfs_design.html.
3 https://spark.apache.org/.
4 https://hadoop.apache.org/docs/current/hadoop-yarn/hadoop-yarn-site/YARN.html.
5 https://zeppelin.apache.org/.
6 https://ambari.apache.org/.
7 https://ranger.apache.org/.
8 https://flume.apache.org.
9 https://sqoop.apache.org/.

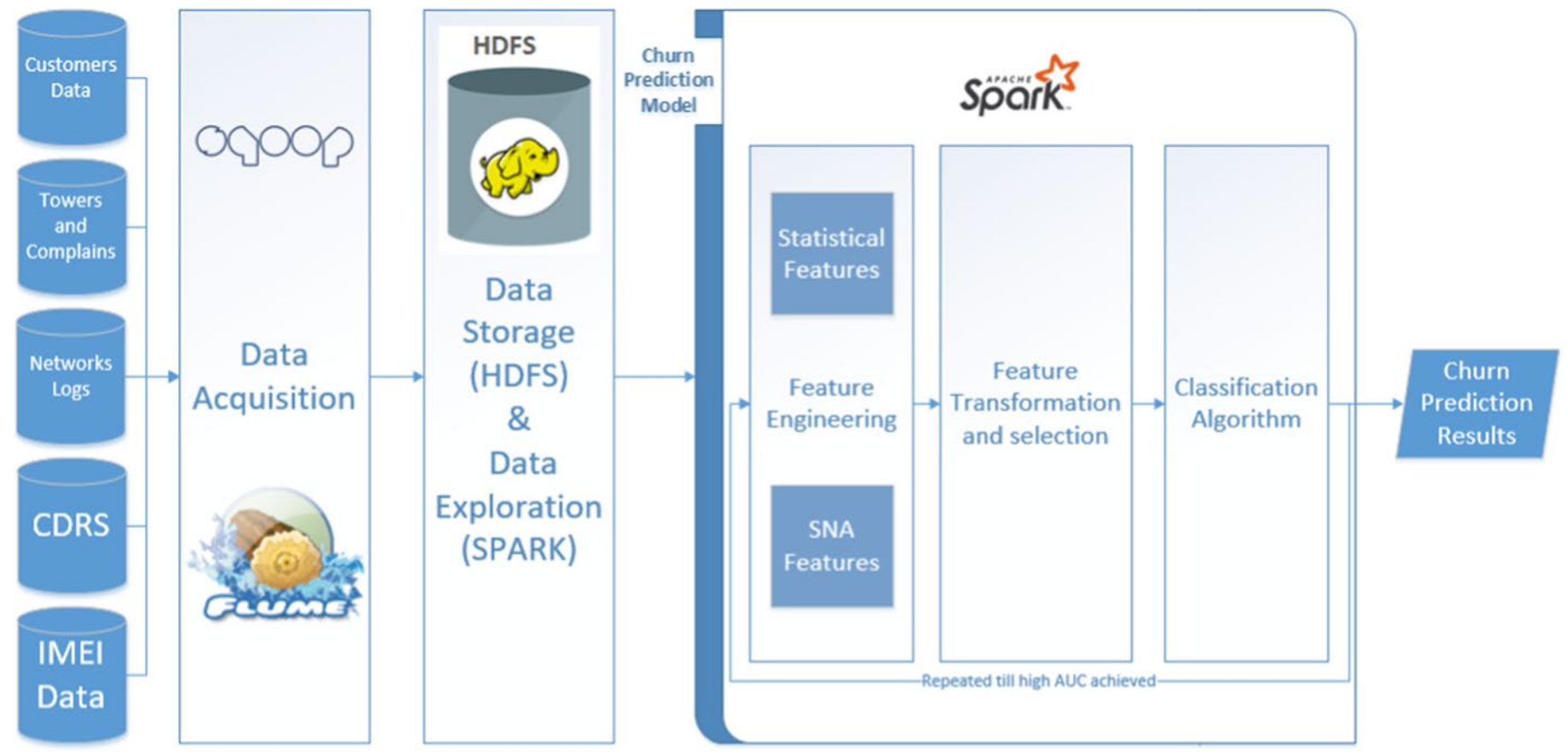

**Fig. 2** Proposed churn Prediction System Architecture

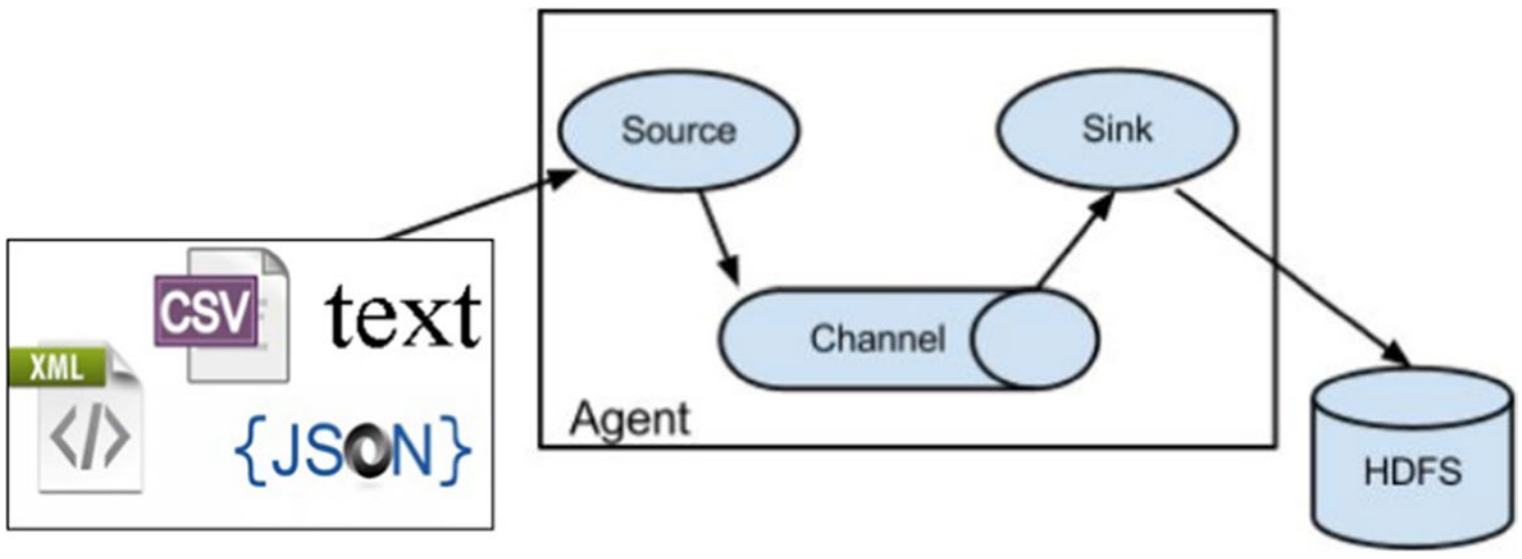

**Fig. 3** Apache Flume configured system architecture

the designed architecture of flume in SYTL-BD. There are three main components in FLUME. These components are the data Source, the Channel where the data moves and the Sink where the data is transported.

Flume agents transporting files exist in the defined Spooling Directory Source using one channel, as configured in SYTL-BD. This channel is defined as Memory Channel because it performed better than the other channels in FLUME. The data moves across the channel to be finally written in the sink which is HDFS. The data transformed to HDFS keep in the same format type as it was.

Apache SQOOP is the distributed tool used to transfer the bulk of data between HDFS and relational databases (Structured data). This tool was used to transfer all the data which exists in databases into HDFS by using Map jobs. Figure 4 shows the architecture of SQOOP import process where four mappers are defined by default. Each Map job selects part of the data and moves it to HDFS. The data is saved in CSV file type after being transported by SQOOP to HDFS.

After transporting all the data from its sources into HDFS, it was important to choose the appropriate file type that gives the best performance in regards to space utilization and execution time. This experiment was done using spark engine where

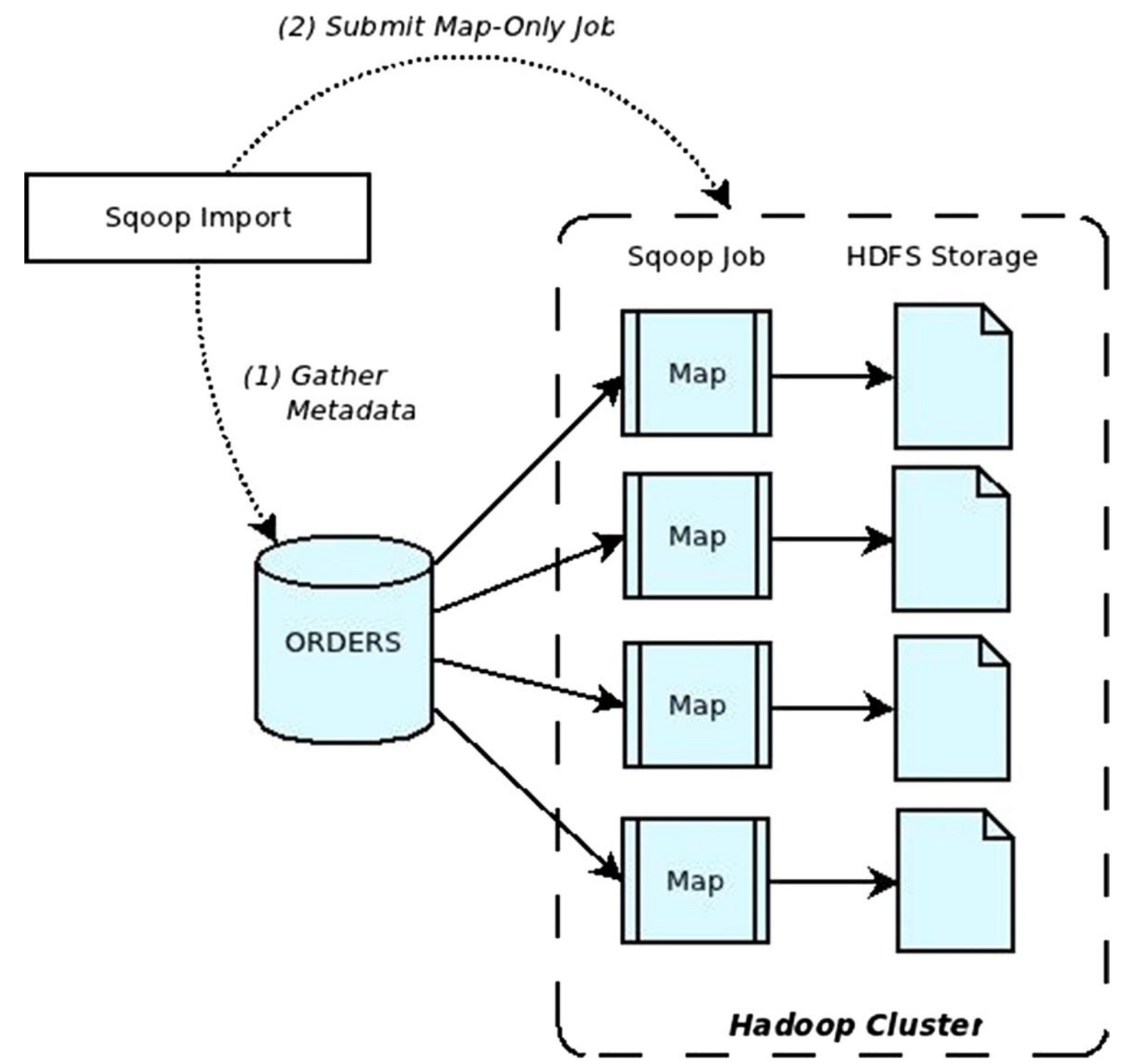

**Fig. 4** Apache SQOOP data import architecture

Data Frame library[10] was used to transform 1 terra byte of CSV data into Apache Parquet[11] file type and Apache Avro[12] file type. In addition to that, three compression scenarios were taken into consideration in this experiment.

Parquet file type was the chosen format type that gave the best results. It is a columnar storage format since it has efficient performance compared with the others, especially in dealing with feature engineering and data exploration tasks. On the other hand, using Parquet file type with Snappy Compression technique gave the best space utilization. Figure 5 shows some comparison between file types.

## Feature engineering

The data was processed to convert it from its raw status into features to be used in machine learning algorithms. This process took the longest time due to the huge numbers of columns. The first idea was to aggregate values of columns per month (average, count, sum, max, min ...) for each numerical column per customer, and the count of distinct values for categorical columns.

Another type of features was calculated based on the social activities of the customers through SMS and calls. Spark engine is used for both statistical and social features, the library used for SNA features is the Graph Frame.

[10] https://spark.apache.org/docs/latest/sql-programming-guide.html.

[11] https://parquet.apache.org/.

[12] https://avro.apache.org/.

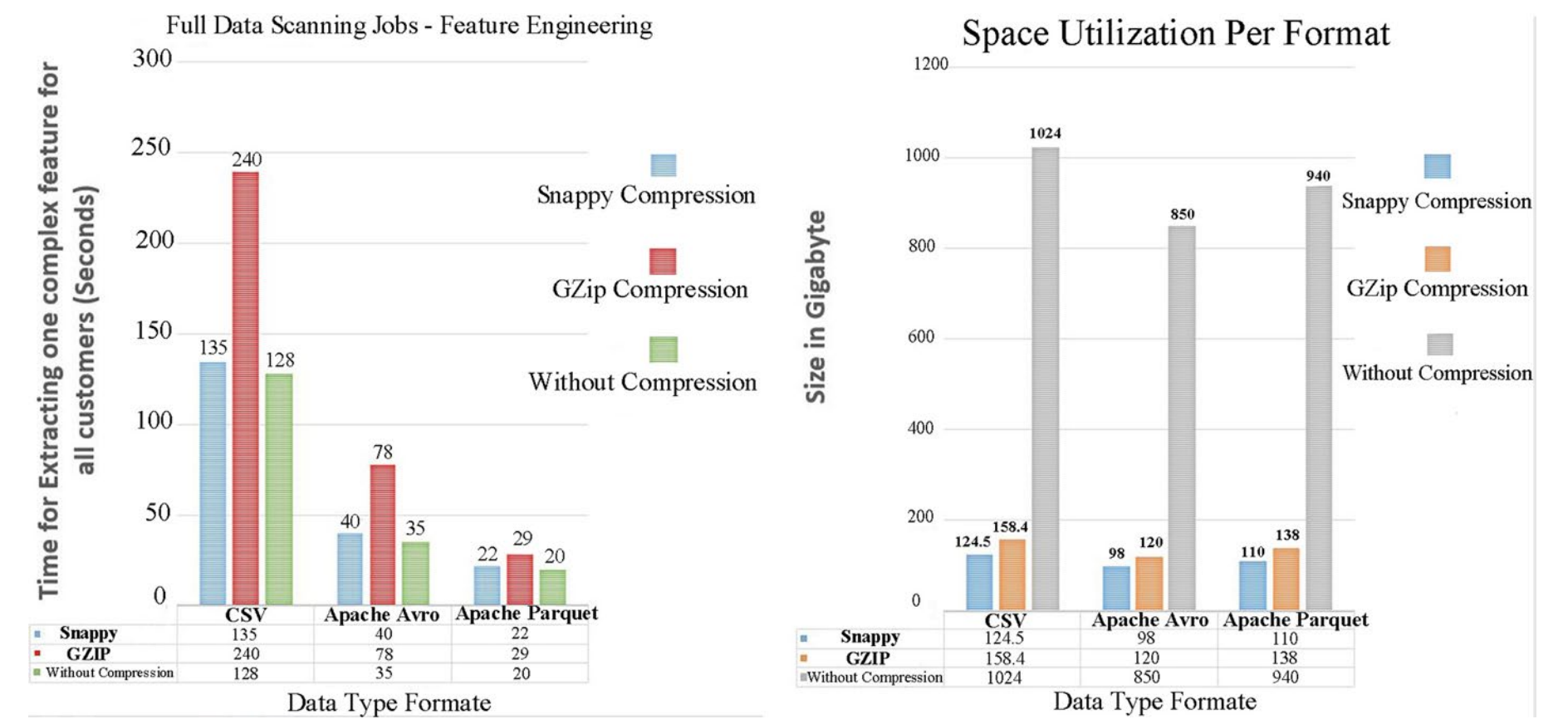

**Fig. 5** Differences in space utilization and execution time per file type

- *Statistics features* These features are generated from all types of CDRs, such as the average of calls made by the customer per month, the average of upload/download internet access, the number of subscribed packages, the percentage of Radio Access Type per site in month, the ratio of calls count on SMS count and many features generated from aggregating data of the CDRs.

Since we have data related to all customers' actions in the network, we aggregated the data related to Calls, SMS, MMS, and internet usage for each customer per day, week, and month for each action during the nine months. Therefore, the number of generated features increased more than three times the number of the columns. In addition, we entered the features related to complaints submitted from the customers from all systems. Some features were related to the number of complaints, the percentage of coverage complaints to the whole complaints submitted, the average duration between each two complaints sequentially, the duration in "Hours" to close the complaint, the closure result, and other features.

The features related to IMEI data such as the type of device, the brand, dual or mono device, and how many devices the customer changed were extracted.

We did many rounds of brainstorming with seniors in the marketing section to decide what features to create in addition to those mentioned in some researches. We created many features like percentage of incoming/out-coming calls, SMS, MMS to the competitors and landlines, binary features to show if customers were subscribing some services or not, rate of internet usage between 2G, 3G and 4G, number of devices used each month, number of days being out of coverage, percentage of friends related to competitor, and hundred of other features.

Figures 6 and 7 visualize some of the basic categorical and numerical features to give more insight on the deference between churn and non-churn classes.

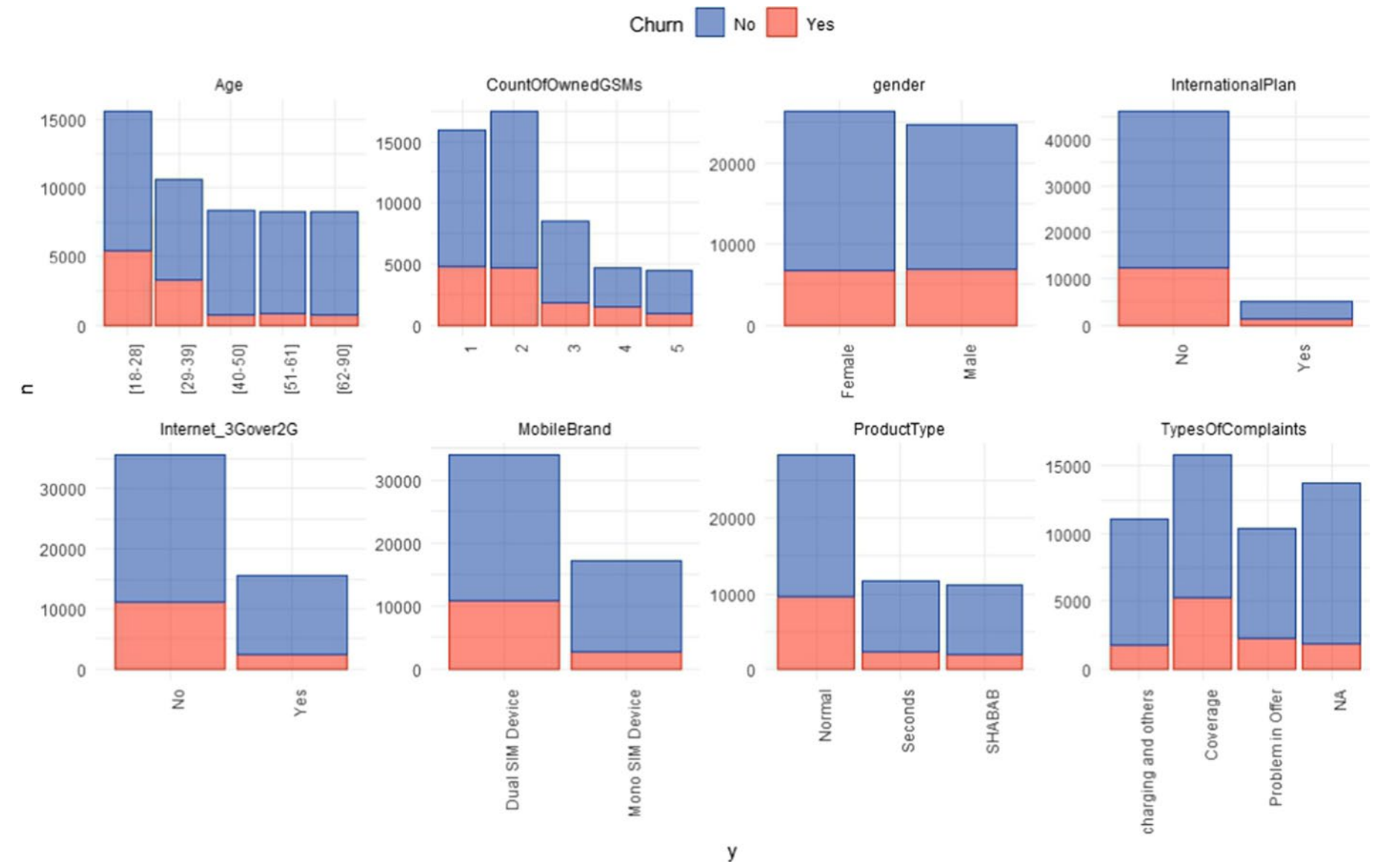

**Fig. 6** Distribution of some main categorical features

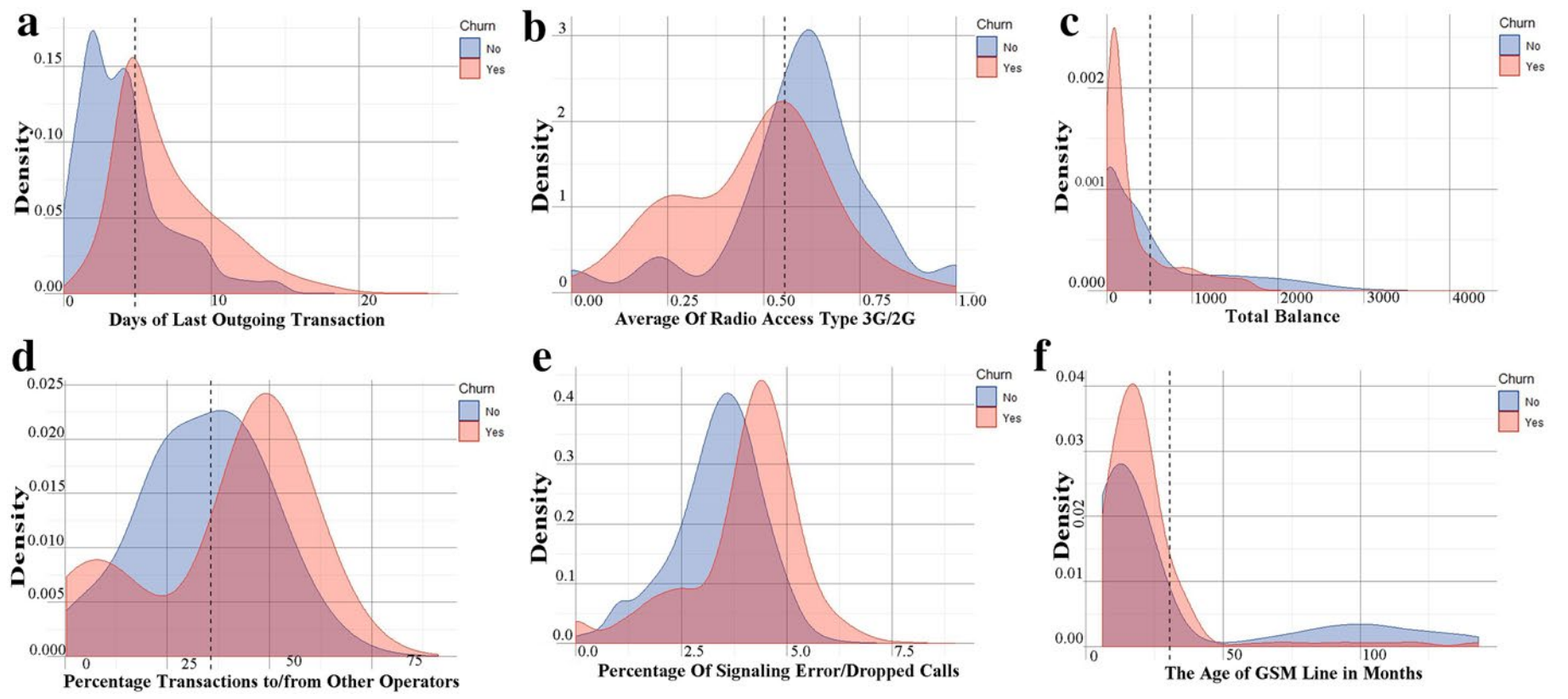

**Fig. 7** Feature distribution for some main numerical features. Panel (**a**) visualizes the distribution of Day of Last Outgoing Transaction feature. Panel (**b**) visualizes the feature distribution of Average Radio Access Type Between 3G and 2G. Panel (**c**) also visualizes the distribution of Total Balance feature. Panel (**d**) shows the feature distribution of Percentage Transaction with other operators. Similarly, panel (**e**) visualizes the distribution of Percentage of Signaling Error/Dropped calls. Finally, panel (**f**) visualizes the distribution of the GSM Age feature. The red color is used in all panels to represent the churned customers' class and the blue one for active customers' class

- *Social Network Analysis features* Data transformation and preparation are performed to summarize the connections between every two customers and build a social network graph based on CDR data taken for the last 4 months. Graph frame library on spark is used to accomplish this work. The social network graph consists of Nodes and edges.

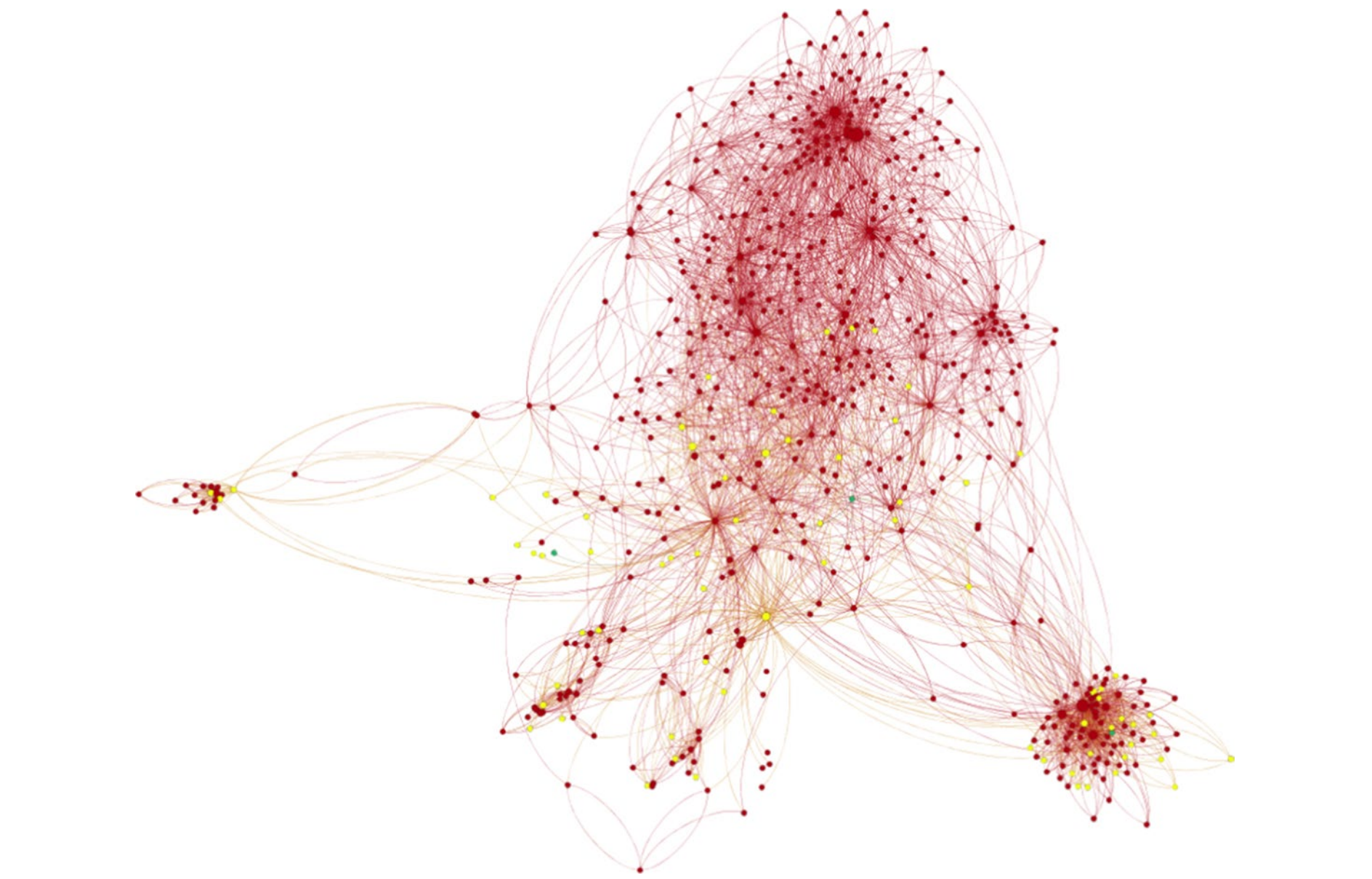

**Fig. 8** Visualization for a sample of the Syrian social community

- Nodes: represent GSM number of subscribers.
- Edges: represent interactions between subscribers (Calls, SMS, and MMS). The graph edges are directed since we have A to B and B to A.

  Figure 8 visualizes a sample of the build social network in SyriaTel where the red nodes are SyriaTel's customers and the Yellow nodes are MTN's Customers, the lines between the nodes express the interaction between the nodes.

  The total social graph contained about 15 million nodes that represent SyriaTel, MTN, and Baseline numbers and more than 2.5 Billion edges.

Graph-based features are extracted from the social graph. The graph is a weighted directed graph. We built three graphs depending on the used edges' weight. The weight of edges is the number of shared events between every two customers. We used three types of weights: (1) the normalized calling duration between customers, (2) the normalized total number of calls, SMS, and MMS, (3) the mean of the previous two normalized weights. The normalization process varies according to the algorithm used to extract the features as we see in the formulas of these algorithms. Based on the directed graphs, we use PageRank [19], Sender Rank [20] algorithms to produce two features for each graph.

- The weighted Page Rank equation is defined as follows

$$PR(m) = (1-d) + d * \sum_{n \in N(m)} \frac{W_{n \to m}}{\sum_{n' \in N(n)} W_{n \to n'}} PR(n) \tag{1}$$

- While the weighted Sender Rank equation is defined as follow

$$SR(m) = (1-d) + d * \sum_{n \in N(m)} \frac{W_{m \to n}}{\sum_{n' \in N(n)} W_{n \to n'}} SR(n) \tag{2}$$

Graph networks related to telecom data may contain two types of nodes. First, nodes with zero outgoing and many incoming interactions. Second, nodes with zero-incoming and many outgoing interactions. These two kinds of nodes are called Sink nodes.

In regards to Eq. (1), the nodes with zero outgoing edges are the Sinks while in Eq. (2) the Sinks are the nodes with zero-incoming edges. The damping factor d is used here to prevent these Sinks from getting higher SR or PR values each round of calculation. Damping factor in telecom social graph is used to represent the interaction-through probability .The first part (1-d) represents the chance to randomly select a sink node while the d is used to make sure that the sum of PageRanks or SenderRanks is equal to 1 at the end. In addition to that, it prevents the nodes with zero-outgoing edges to get zero SenderRank values and the nodes with zero-incoming edges to get zero PageRank values since these values will be passed to the sink nodes each round. If d =1, the equations need an infinite number of iterations to reach convergence. While a low d value will make the calculations easier but will give incorrect results. We assumed to set the d value to be 0.85 as mentioned in most of the research [21, 22].

N(m) is the list of friends for the customer (m) in his social network. $W_{n \to m}$ is the directed edge weight from n to m. $\frac{W_{n \to m}}{\sum_{n' \in N(n)} W_{n \to n'}}$ is the normalized weight of the directed edge from n to m. The same description is used for sender rank.

Due to the random walk nature of the Eqs. (1) and (2), PR and SR will be stable after a number of iterations. These values indicate the importance of the customers since the higher values of PR(m) and SR(m) corresponds to the higher importance of customers in the social network.

Other SNA features like the degree of centrality, IN and OUT degree which is the number of distinct friends in receive and send behavior were calculated.

The feature Neighbor Connectivity based on degree centrality which means the average connectivity of neighbors for each customer is also calculated [23].

- Neighbor Connectivity equation is defined as follow

$$NC(m) = \frac{\sum_{k \in N(m)} |N(k)|}{|N(m)|} \tag{3}$$

The local clustering coefficient for each customer is also calculated. This feature tells us how close the customer's friends are (number of existing connections in a neighborhood divided by the number of all possible connections) [24].

- local clustering coefficient equation is defined as follow

$$LC(m) = \sum_{k \in N(m)} \frac{|N(m) \cap N(k)|}{|N(m)| * (|N(m)| - 1)} \tag{4}$$

This social network is also used to find similar customers in the network based on mutual friend concept. Each customer has 2 similarity features with the other customers in his network, like Jaccard similarity, and Cosine similarity. These calculations were done for each distinct couple in the social network, where each customer will have two calculations in the network. To reduce this complexity, customers who don't have mutual friends are excluded from these calculations. The highest values for both measures are selected for each customer ( top Jaccard and Cosine similarity for similar SyriaTel customer and top Jaccard and Cosine similarity for similar MTN customer). Jaccard measure: normalize the number of mutual friends based on the union of the both friends lists, [25].

- Jaccard similarity equation between customer(m) and customer(k) is defined as follows:

$$JS(m,k) = \frac{\left|N(m) \cap N(k)\right|}{\left|N(m) \cup N(k)\right|} \tag{5}$$

Another similarity measure is the Cosine measure which is similar to Jaccard's. On the other hand, this similarity measure calculates the Cosine of the angle between every two customers' vectors where the vector is the friend list of each customer [25].

- Cosine similarity equation between customer(m) and customer(k) is defined as follows:

$$JS(m,k) = \frac{\left|N(m) \cap N(k)\right|}{\sqrt{|N(m)|\left|N(k)\right|}} \tag{6}$$

The cosign similarity is useful when the customer is in the phase of leaving the company to the competitor, where he starts building his network on the new GSM line to be similar to the old being churned, taking into consideration that the new line has a small friends list compared with the old one.

These features are used for the first time to enhance the prediction of churn, and they have a positive effect along with the other statistical features. The distribution of the main SNA features are presented in Fig. 9.

Table 1 shows some calculated main SNA features with illustration.

### Features transformation and selection

Some features such as Contract ID, MSISDN and other unique features for all customers were removed. They are not used in the training process because they have a direct correlation with the target output (specific to the customer itself). We deleted features with identical values or missing values, deleted duplicated features, and features that have few numeric values. We found that more than half of the features have more than 98% of missing values. We tried to delete all features that have at least one null value, but this method gave bad results.

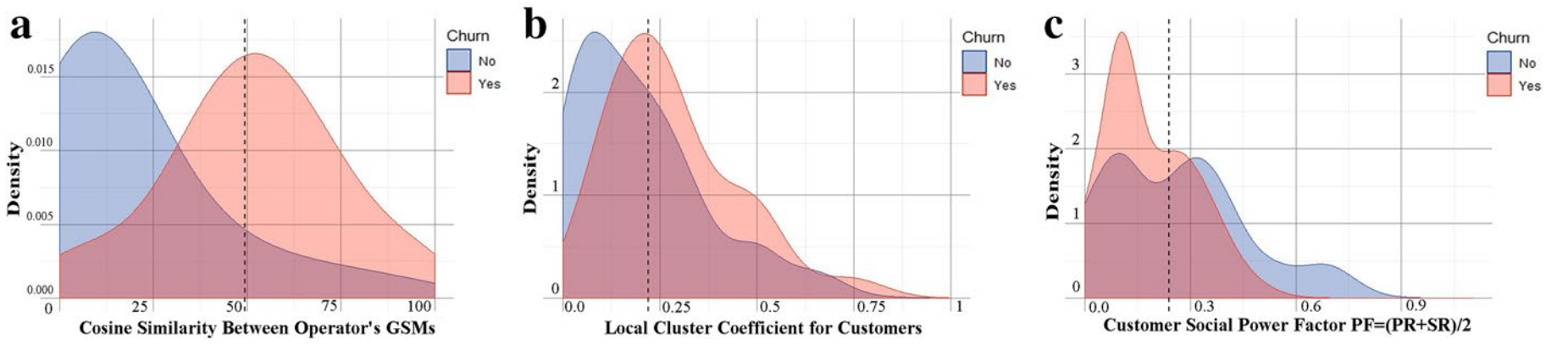

**Fig. 9** Distribution of some main SNA features, panel (**a**) visualizes the feature distribution of Cosine Similarity Between GSM Operators, panel (**b**) visualizes the distribution of Local Cluster Coefficient feature, and panel (**c**) visualizes the distribution of Social Power Factor feature. The red color is used in all panels to represent the churned customers' class and the blue one for active customers' class

**Table 1 Some main SNA features with description**

| Feature name | Feature description |
|---|---|
| In-Degree | Number of friends connecting with the customer |
| Out-Degree | Number of friends the customer connecting with |
| Max-Cosine-Sim-MTN | Maximum Cosine similarity with other operators' customers |
| Max-Cosine-Sim-SyriaTel | Maximum Cosine similarity with SyriaTel customers |
| Max-Jaccard-SIM-MTN | Maximum Jaccard similarity with other operators' customers |
| Max-Jaccard-Sim-SyriaTel | Maximum Jaccard similarity with SyriaTel customers |
| SR | Weighted Sender Rank in social graph |
| PR | Weighted Page Rank in social graph |
| PF "Social Power Factor" | Average of weighted Page Rank and Sender Rank in social graph |
| Betweenness | # of short paths between any two people in the social network passes through this customer node |
| LLC "Local Cluster Coefficient" | How much the customer friends know each other |
| NC "Neighborhood Connectivity" | The number of friends and friends of friends for the customer |

Finally, we filled out the missing values with other values derived from either the same features or other features. This method is preferable so that it enables us to use the information in most features for the training process. We applied the following:

- Records that contain more than 90% of missing features were deleted.
- Features that have more than 70% of missing values were deleted.
- For the missing categories in categorical features, they were replaced by a new category called 'Other'.
- The missing numerical values were replaced with the average of the feature.
- The number of categorical features were 78, the first 31 most frequent categories were chosen and the remaining categories were replaced with a new category, so the total number is 32 categories.
- There are some other features with a numeric character but they contain only a limited number of duplicate values in more than one record. This indicates that they are categorical so we have dealt with them as categorical features, but the experiment shows that they perform worse with the model, so that they have been deleted.

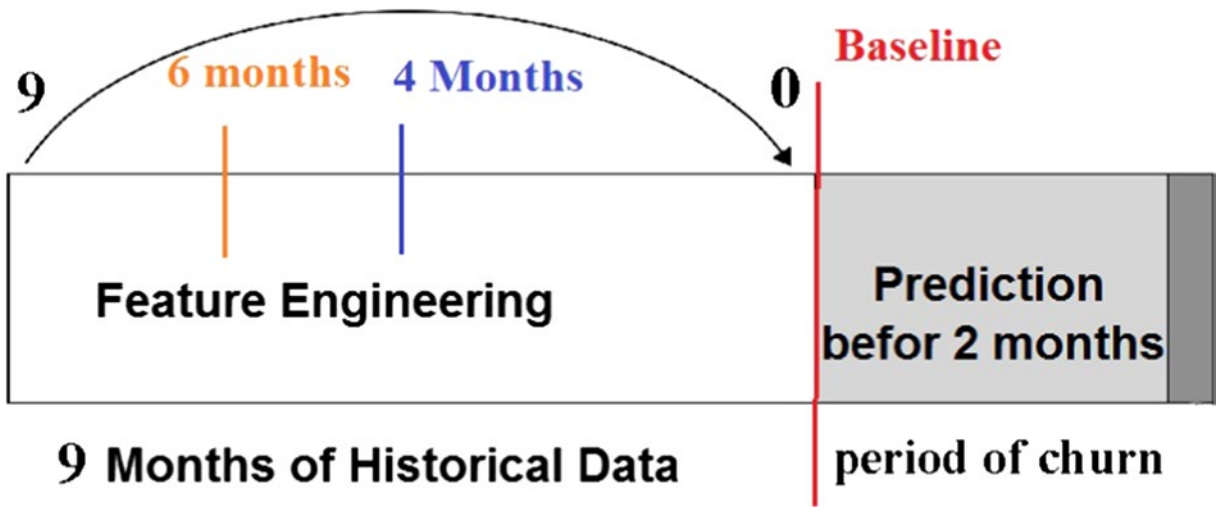

**Fig. 10** Periods of historical and future data

We have also calculated the correlation between numerical features using Pearson and removed the correlated features. This removal had no effect on the final result. Many other methods were tested, but this applied approach gave the best performance of the four algorithms. The number of features after this operation exceeded 2000 features at the end.

We need this data labeled for training and testing, we contacted experts from the marketing section to provide us with labeled sample of GSM, so they provide us with a prepaid customers in idle phase after 2 months of the nine months data, considering them as churners. The other non-churned customers were labeled as Active customers (customers acquired in the last 4 months are excluded). The total count of the sample where 5 million customers containing 300,000 churned customers and 4,700,000 active customers. Figure 10 shows the periods of historical data and the future period when the customer may leave the company.

The experts in marketing decided to predict the churn before 2 months of the actual churn action, in order to have sufficient time for proactive action with these customers.

## Classification

The solution we proposed divided the data into two groups: the training group and the testing group. The training group consists of 70% of the dataset and aims to train the algorithms. The test group contains 30% of the dataset and is used to test the algorithms. The hyperparameters of the algorithms were optimized using K-fold cross-validation. The value of k was 10. The target class is unbalanced, and this could cause a significant negative impact on the final models. We dealt with this problem in our research by rebalancing the sample of training by taking a sample of data to make the two classes balanced [25]. We started with oversampling by duplicating the churn class to be balanced with the other class. We also used the random undersampling method, which reduces the sample size of the large class to become balanced with the second class. This method is the same as the one used in more than one research papers [8, 26]. It gave the best result for some algorithms. The training sample size became 420,000.

We started training Decision Tree algorithm and optimizing the depth and the maximum number of nodes hyperparameters. We experimented with several values, the optimized number of nodes was 398 nodes in the tree and the depth value was 20.

Random Forest algorithm was also trained, we optimized the number of trees hyper-parameter. We experimented with building the model by changing the values of this parameter every time in 100, 200, 300, 400 and 500 trees. The best results show that the best number of trees was 200 trees. Increasing the number of trees after 200 will not give a significant increase in the performance. GBM algorithm was trained and tested on the same data, we optimized the number of trees hyper-parameter with values up to 500 trees. The best value after the experiment was also 200 trees. GBM gave better results than RF and DT. We finally installed XGBOOST on spark 2.3 framework and integrated it with ML library in spark and applied the same steps with the past three algorithms. We also optimized the number of trees, and the best value after multiple experiments was 180 trees.

## Results and discussion

The results were analyzed to compare the performance regarding the different sizes of training data. Dealing with unbalanced dataset using the three scenarios were also analyzed. The first main concern was about choosing the appropriate sliding window for data to extract statistical and SNA features. How much historical data is needed in features engineering phase?

In Fig. 11, M1 refer to the first month before the baseline and M9 refer to the ninth month before baseline. The features of month N are aggregated from the N-month sliding data window (from month 1 to month N). As Fig. 11a presents, we can confirm that increasing the volume of training data to get statistical features increases the performance of the classification algorithms. However, the addition of the oldest three months did not provide any enhancement on model performance. When only using statistical features, the highest value of AUC reached 84%.

The Social Network Analysis features had a different scenario, when the best sliding window to build the social graph and extract appropriate SNA features was during the last four months before the baseline, as shown in Fig. 11b. Adding more old data will adversely affect the performance of the model. The highest AUC value reached by using only SNA features was 75.3%.

Depending on the above two different scenarios, the last 6 months of the raw dataset was used to extract the statistical features, while the last four months of that dataset was only used to extract the SNA features. Figure 10 presents the best sliding window to extract SNA features in orange and the blue one is for statistical features while the red line represents the baseline.

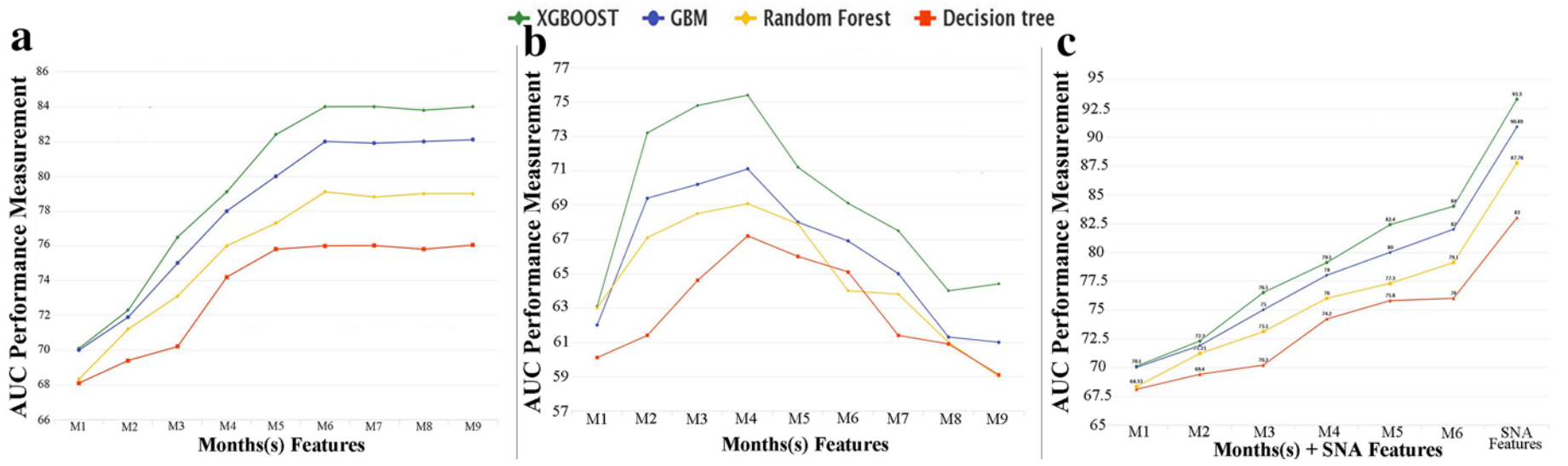

**Fig. 11** Performance of classification algorithms per sliding window and feature type. Panel (**a**) shows the improvement of churn predictive model using Statistical Features related to different historical periods, panel (**b**) presents the changes in predictive model improvement using SNA Features related to the same historical periods, and panel (**c**) presents the enhancement of churn predictive model when using both statistical and SNA Features

**Table 2 Comparing AUC results before and after adding SNA to statistical features**

| Features | XGBOOST (%) | GSM (B) (%) | Random Forest (%) | Decision Tree (%) |
|---|---|---|---|---|
| Statistical features | 84 | 82 | 79.1 | 76 |
| SNA features | 75.3 | 71 | 69 | 67.2 |
| Statistical and SNA features | 93.3 | 90.89 | 87.76 | 83 |

By adding SNA features with the statistical features to the classification algorithms, the results increased significantly. As presented in Table 2 and Fig. 11c, the addition of both types of features made a good enhancement to the performance of the churn predictive model, where the max reached value of AUC was 93.3%.

The second concern taken into consideration was the problem of the unbalanced dataset since three experiments were applied for all classification algorithms. These experiments are: (1) classification with undersampling technique, (2) classification with oversampling technique, (3) classification without balancing the dataset. Table 3 shows that both XGBOOST and GBM algorithms gave the best performance without any rebalancing techniques, while Random Forest and Decision Tree algorithms gave a higher performance by using undersampling techniques.

As displayed in Fig. 11 and depending on Tables 2 and 3, we confirm that XGBOOST algorithm outperformed the rest of the tested algorithms with an AUC value of 93.3% so that it has been chosen to be the classification algorithm in this proposed predictive model. GBM algorithm occupied second place with an AUC value of 90.89% while Random Forest and Decision Trees came last in AUC ranking with values of 87.76% and 83% sequentially. Figure 12 shows the ROC curves for the four algorithms.

The top important features that contribute to predict the churn were ranked using Gain measure [27]. The high gain value of the feature means the more important it is in predicting the churn. The important features according to XGBOOST algorithm are presented in Fig. 13 before and after merging SNA and statistical features.

As presented in Fig. 13, adding the Social Network Analysis features changed the ranking of the important features. The MTN Cosine similarity was the most important feature since the customers with higher MTN Cosine similarity are more likely to churn

**Table 3 Comparing the AUC results of Machine Learning algorithms with each balancing technique**

| Technique used for unbalanced dataset | XGBOOST (%) | GSM (B) (%) | Random Forest (%) | Decision Tree (%) |
|---|---|---|---|---|
| Oversampling | 92 | 90.01 | 84.2 | 76.25 |
| Undersampling | 93.12 | 90.21 | *87.76* | *83* |
| Without balancing | *93.3* | *90.89* | 78.47 | 72.2 |

As shown in the table each Machine Learning algorithm experimented with three different scenarios with regards to the problem of the unbalanced dataset. The best results of AUC presented in italics indicate the best technique that fits each algorithm

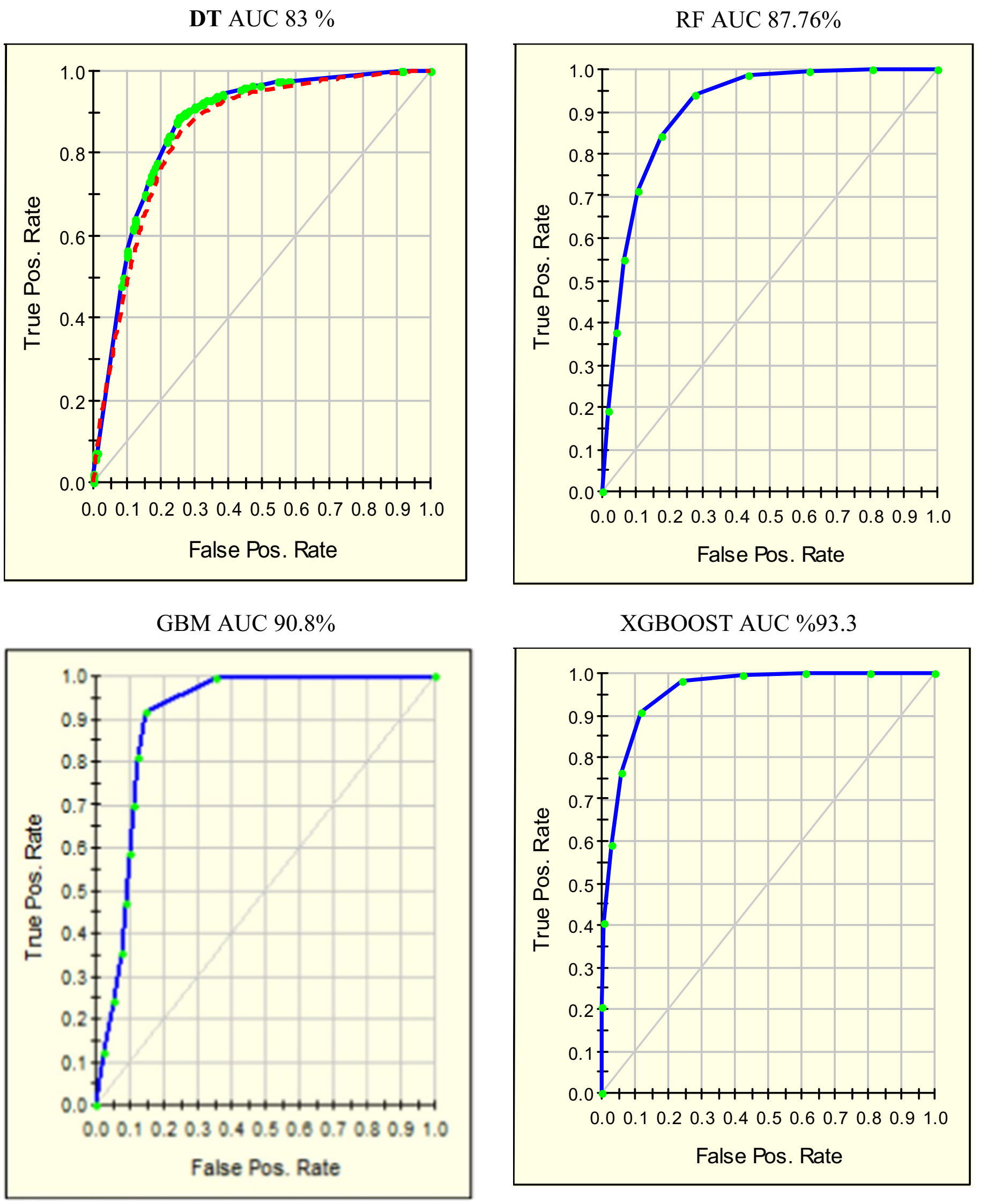

**Fig. 12** Receiver operating characteristic curve for each classification algorithm

regardless of the other features like balance, internet usage, and in/out calls. Figure 9a displays the distribution of this feature

By analyzing this feature, most of the customers generally stayed active for a period of time before terminating or stopping the use of their GSMs. This case probably happens because the customer needs to make sure that most of his important incoming calls and contacts have moved to the new line. In other words, the customer could wait for a period of time to make sure that most of his important people have known the new GSM number. This case also could be justified as the customer need to finish the remaining balance in the GSM before he stops using it. Figure 14 shows an example of a tracked churned customer. this figure presents the phases of moving his community to the other operator's GSM.

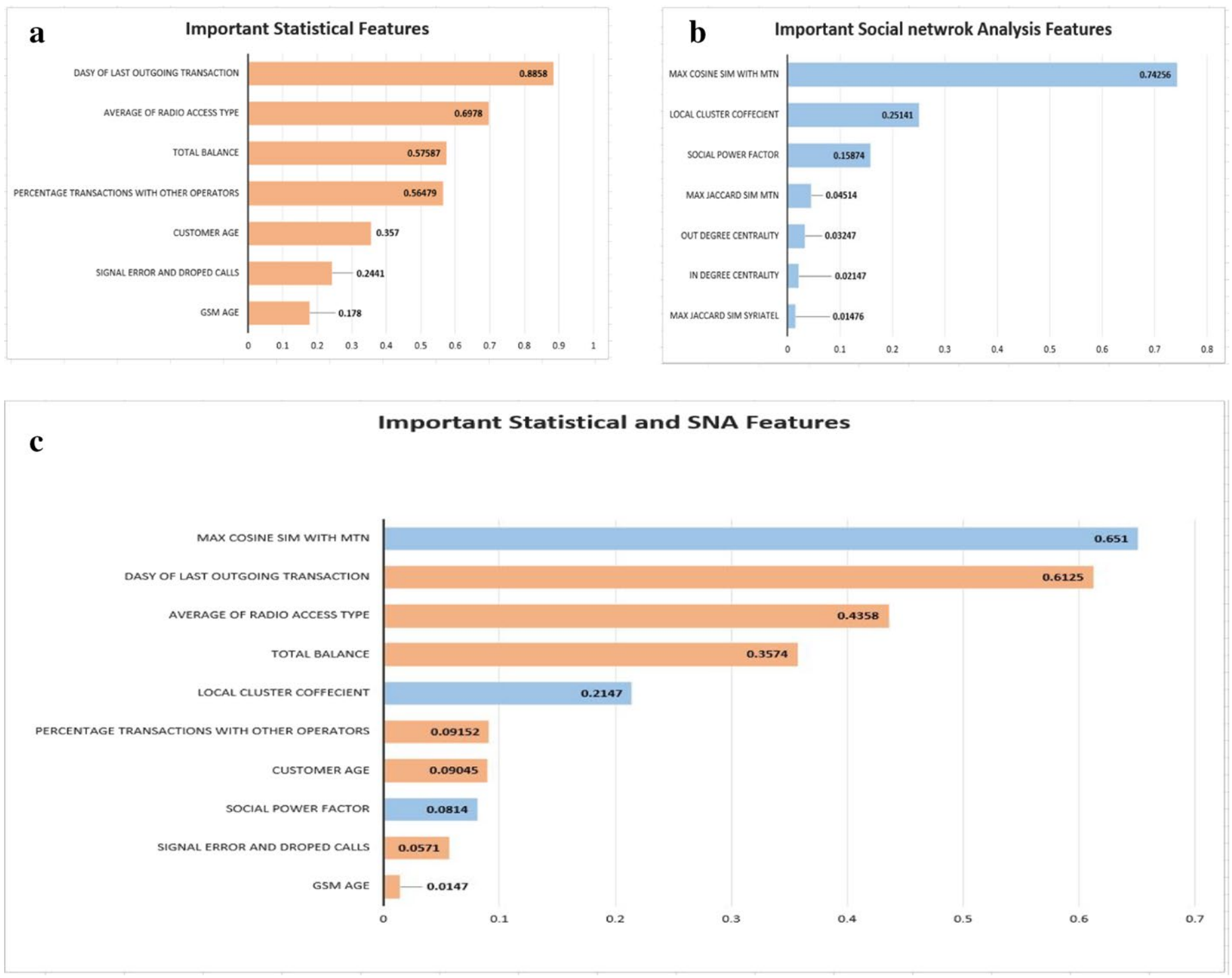

**Fig. 13** Important features per type according to XGBOOST, panel (**a**) shows the ranking of important Statistical features, panel (**b**) shows the ranking of important Social Network Analysis features, while panel (**c**) shows how adding both Statistical and SNA features re-ranks their importance in XGBOOST algorithm. The orange color is used in all panels to represent the Statistical features and the blue one for SNA features

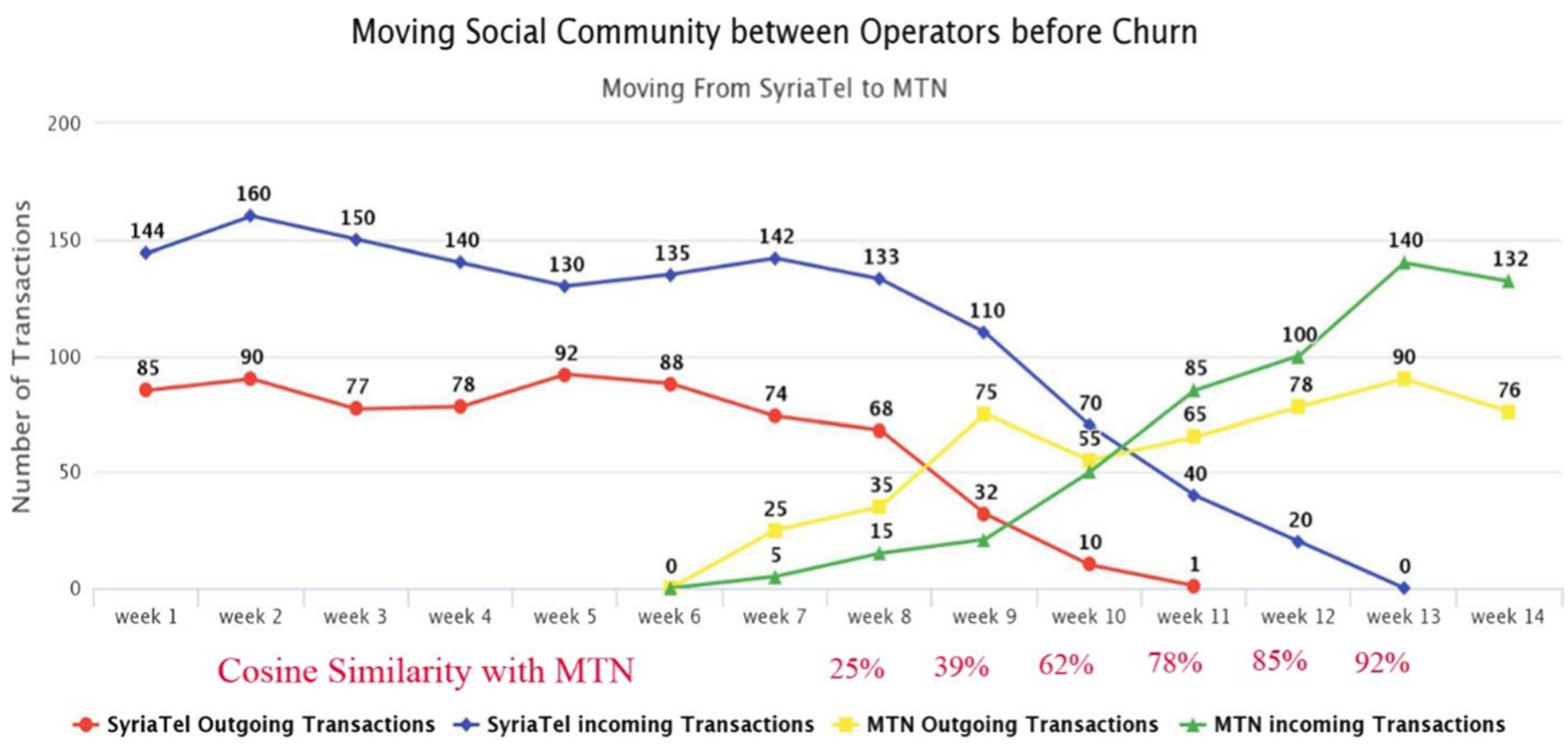

**Fig. 14** Tracked churned customer with Max Cosine MTN Similarity values per week

The customer bought GSM from the competitor in week 7 and terminated SyriaTel's GSM in week 14 before being out of coverage in week 13 and week 14. The result of the Cosine similarity is also displayed in the same figure.

The second important feature is Days of Last Outgoing transaction. As shown in Fig. 7a, most churners stay longer period than non-churners without making any transaction.

The third important feature is total balance since most churners had low balance compared with the active customers regardless of the reason of churn, Fig. 7c shows the distribution of total balance feature for churners and non-churners customers.

The fourth feature in importance is Average of Radio Access Type where most of the churners had more 2G internet sessions than 3G sessions, as the speed and quality of 2G internet sessions is much less than these of 3G sessions. Figure 7b shows the distribution of this feature where the Average RAT is lower for most of the churners compared with that of non-churners. The customers are more likely to churn if they are heavy internet users and there is a better 3G coverage provided by the competitor. By analyzing this feature, 68% of churners are internet users, 65% of them have low Average Radio Access Type value.

Local Cluster Coefficient is another SNA feature, it's ranked fifth in importance to predict the churn since the customers with very low LCC value as shown in Fig. 9b are less likely to churn. This could be justified because some customers are using these personal GSMs for business objectives. They need to preserve their numbers in order not to lose any of their customers. Most of these customers have more than two GSMs. They communicate with lots of people, most of these people don't know each other (there is no interaction between them). A sample of customers with very low LCC were contacted to check this case. The results show that most of them were related to Cafes, Restaurants, Shaving shops, Hairdressers, Libraries, Game Shops, Medical clinics, and others.

The sixth important feature is the Percentage of Transactions to/from other Operator, this value becomes bigger for churners. The explanation here relies on the effect of friends on the churn decision, since the affiliation of most of customer's friends to the other operator may be evidence of the good reputation or the strong existence of the competing company in that region or community. Therefore, this can result in the customer being influenced by the surrounding environment, so he moves to the competing company. The higher value of this feature may increase the likelihood of churn, Fig. 7d displays the distribution of this feature.

Other features like Customer age is also ranked at the seventh place in importance since the customers who are less than 32 years old have more likelihood to churn than the others. This can be explained by the fact that young people are always looking for the best to meet their needs in better, higher quality, and less expensive services as the volume of communication, the use of Internet, and other services are much higher compared to services of customers of different ages. Figure 6 shows the distribution of this feature regarding the churn class. The social power factor feature is the third SNA feature that is considered one of the top important features to predict the churn. As presented in Fig. 9c the higher power factor value means the less likely to churn. As also shown in Fig. 7e, the customers with high Signal Errors and dropped calls are most likely to churn.

Depending on what was mentioned previously and as shown in Figs. 11c, 12c, we belive that Social Network Analysis features have a good contribution to increase the performance of churn prediction model, since they gave a different insight to the customer from the social point of view.

**Table 4 AUC results for classification algorithms on "NotOffered" dataset**

| Algorithm | XGBOOST | GSM (B) | Random Forest | Decision Tree |
|---|---|---|---|---|
| SyriaTel New Data "NotOffered" | 89% | 85.5% | 83.4% | 79.1% |

- *System evaluation* We evaluated the system by using new up to date dataset. The test was conducted on all prepaid SyriaTel customers without any exception. The population was 7.5 million customers without knowing what their status will be after 2 months. The same models were tested on this data set after being processed as mentioned previously. The dataset for customers who are most likely predicted to churn, was divided into two datasets (Offered, NotOffered). Marketing experts make a proactive action to retain the customers who are predicted to leave SyriaTel from the offered dataset, and the other dataset "NotOffered" left without any action. The results of the test were compared with the customer's status after two months for the two datasets. The results were very good and the best AUC value was 89% for XGBOOST on "NotOffered" and most of the cases were predicted right. Table 4 shows AUC results for the four algorithms on the "NotOffered" dataset.

  The percentage of the retained customers from Offered dataset was about 47% from all customers predicted to churn. In other words, about half of the customers changed their mind regarding churn decision when they got a good offer. This result was very good for the company, increased the revenue and decreased the churn rate by about 1.5%.

## Conclusion

The importance of this type of research in the telecom market is to help companies make more profit. It has become known that predicting churn is one of the most important sources of income to telecom companies. Hence, this research aimed to build a system that predicts the churn of customers in SyriaTel telecom company. These prediction models need to achieve high AUC values. To test and train the model, the sample data is divided into 70% for training and 30% for testing. We chose to perform cross-validation with 10-folds for validation and hyperparameter optimization. We have applied feature engineering, effective feature transformation and selection approach to make the features ready for machine learning algorithms. In addition, we encountered another problem: the data was not balanced. Only about 5% of the entries represent customers' churn. This problem was solved by undersampling or using trees algorithms not affected by this problem. Four tree based algorithms were chosen because of their diversity and applicability in this type of prediction. These algorithms are Decision Tree, Random Forest, GBM tree algorithm, and XGBOOST algorithm. The method of preparation and selection of features and entering the mobile social network features had the biggest impact on the success of this model, since the value of AUC in SyriaTel reached 93.301%. XGBOOST tree model achieved the best results in all measurements. The AUC value was 93.301%. The GBM algorithm comes in the second place and the random forest and Decision Tree came third and fourth regarding AUC values. We have evaluated the models by fitting a new dataset related to different periods and without any proactive action from marketing, XGBOOST also gave the best result with 89% AUC. The decrease in result could be due to the non-stationary data model phenomenon, so the model needs training each period of time.

The use of the Social Network Analysis features enhance the results of predicting the churn in telecom.

**Abbreviations**

CDR: call detail record; CRM: customer relationship management; SMS: short message service; HDFS: Hadoop Distributed File System; XGBoost: Extreme Gradient Boosting; RF: Random Forest; DT: Decision Tree; AUC: Area Under the Curve; GSM: global system for mobile communications; IMEI: International Mobile Equipment Identity; SNA: Social Network Analysis; GBM: Gradient Boosted Machine; ROI: return on investment; ROC: receiver operating characteristic; PCA: principal component analysis; RAM: random access memory; MMS: multimedia messaging service; CSV: comma-separated values; JSON: javascript object notation; XML: extensible markup language.

**Authors' contributions**

AKA took the role of performing the literature review, building the big data platform, working on the proposed churn model. he conducted the experiments and wrote the manuscript. AJ and KJ took on a supervisory role and oversaw the completion of the work. All authors read and approved the final manuscript.

**Acknowledgements**

Many thanks to SyriaTel, Mrs. CEO Majda Sakr, Mr. Murid Atassi, and Mr. Adham Troudi for support and motivation. Thanks for Mr. Mhd Assaf, Mr. Nour Almulhem, Mr.william Soulaiman, Mr. Ammar Asaad, Mr. Soulaiman Moualla, Mr. Ahmad Ali, and Miss. Marwa Hanhoun for their co-operation and help. Thanks to Mr. Kasem Jamil Ahmad and Mr. Fahmi Ammareen for reviewing the final version of this paper.

**Competing interests**

The authors declare that they have no competing interests.

**Availability of data and materials**

The data is not available to public because of the restriction applied on it from SyriaTel Telcom company, since the license was granted for this study. The data is available to researchers in SyriaTel Company and will be available for others after getting the permission from the company.

**Consent for publication**

The authors consent for publication.

**Ethics approval and consent to participate**

All authors give ethics approval and consent to participate in submission and review process.

**Funding**

The authors declare that they have no funding.

## Publisher's Note

Springer Nature remains neutral with regard to jurisdictional claims in published maps and institutional affiliations.